\documentclass[pre,twocolumn,amsmath,amssymb,floatfix]{revtex4}

\usepackage{graphicx}
\usepackage{dcolumn} 
\usepackage{bm}      
\usepackage{color}

\begin{document}

\title{Helicity cascades in rotating turbulence}
\author{P.D. Mininni$^{1,2}$ and A. Pouquet$^2$}
\affiliation{$^1$Departamento de F\'\i sica, Facultad de Ciencias Exactas y
         Naturales, Universidad de Buenos Aires, Ciudad Universitaria, 1428
         Buenos Aires, Argentina. \\
             $^2$NCAR, P.O. Box 3000, Boulder, Colorado 80307-3000, U.S.A.}
\date{\today}

\begin{abstract}
The effect of helicity (velocity-vorticity correlations) is studied in 
direct numerical simulations of rotating turbulence down to Rossby numbers 
of 0.02. The results suggest that the presence of net helicity plays an 
important role in the dynamics of the flow. In particular, at small Rossby 
number, the energy cascades to large scales, as expected, but helicity then 
can dominate the cascade to small scales. A phenomenological interpretation 
in terms of a direct cascade of helicity slowed down by wave-eddy 
interactions leads to the prediction of new inertial indices for the 
small-scale energy and helicity spectra.
\end{abstract}
\maketitle

\section{Introduction}
Invariants of the equations of motion play an essential role in the behavior 
of turbulent flows. The well-known cascade of energy to the small scales in 
three dimensional hydrodynamic turbulence, associated with the energy 
invariant, has been studied at length since the celebrated paper of 
Kolmogorov \cite{Kolmogorov41}. Less well understood is the role played by 
the second quadratic (but non positive definite) invariant of the Euler 
equations, namely the helicity which embodies the global correlations between 
the velocity field ${\bf u}$ and the vorticity 
\mbox{\boldmath $\omega=\nabla \times {\bf u}$}. Helicity itself plays no 
role in the Kolmogorov (K41) theory of turbulence \cite{Kolmogorov41}. 
Shortly after the discovery that helicity is a quadratic invariant 
of the three-dimensional Euler equation \cite{Moffatt69} (see also 
\cite{Betchov61}), two scenarios were put forward for its dynamical behavior 
\cite{Brissaud73}: a dual cascade of energy and helicity towards smaller 
scales, and a pure helicity cascade with no cascade of energy. Studies of 
absolute equilibrium ensembles for isotropic helical turbulence 
\cite{Kraichnan73} gave support to the former scenario, a result that was 
later confirmed by two-point closure models of turbulence \cite{Andre77} as 
well as by direct numerical simulations (DNS) 
\cite{Borue97,Chen03,Chen03b,Gomez04,Mininni06}.

In non-rotating helical hydrodynamic turbulence, both the helicity and the 
energy cascade towards smaller scales with constant fluxes. The assumption 
that the transfer rates are determined by the energy flux alone gives 
Kolmogorov scaling in the inertial range of both quantities, as is 
observed in the numerical simulations. As a result, the presence of 
helicity may globally arrest the energy transfer (when ${\bf u}$ is 
strictly parallel to \mbox{\boldmath $\omega$}, the nonlinear term -- 
expressed in terms of the Lamb vector 
${\bf u} \times$ \mbox{\boldmath $\omega$} --  is 
zero to within a pressure term), but the energy cascade scaling does not 
differ from that of non-helical turbulence. 

In rotating turbulence, helicity is still an inviscid quadratic invariant. 
Perhaps because of the existence of dual direct cascades in non-rotating 
turbulence, not much attention has been paid in the literature to the 
scaling of net helicity in the rotating case. Helical-wave decompositions 
were introduced in \cite{Cambon89,Waleffe93} (see also 
\cite{Craya58,Herring74}) and were found useful in the study of rotating 
turbulence \cite{Cambon97,Smith99}. Theoretical predictions for the 
helicity spectrum in the presence of strong rotation were also given 
in \cite{Galtier03} in the framework of weak turbulence, under 
the assumption of a dual cascade. Recently, the effect of helicity in 
free-decaying rotating turbulence was studied in numerical simulations 
\cite{Morinishi01}. It was observed that both effects inhibit the energy 
transfer through different mechanisms: helicity diminishes nonlinear 
interactions globally, whereas rotation concentrates nonlinear interactions 
to resonant triads of inertial waves.

The lack of detailed studies of rotating helical flows is remarkable 
considering the relevance of helicity in certain atmospheric processes 
\cite{Lilly88,Kerr96,Markowski98}, such as rotating convective (supercell) 
thunderstorms the predictability of which may be enhanced because of the 
increased stability associated to the weakening of the nonlinear terms.

Recently, high resolution numerical simulations of rotating flows with 
non-helical forcing \cite{Mininni08} showed that, while the velocity and 
vorticity in real space develop anisotropies and large-scale column-like 
structures as expected, the spatial distribution of helicity is more 
homogeneous and isotropic and tends to have a short correlation length. 
This observation motivates the present study. We present results from 
DNS of rotating turbulent flows with helical forcing. The results suggest 
that rotating helical flows behave in a different way than rotating 
non-helical flows. In particular, an inverse cascade of energy and a 
direct cascade of energy and helicity are discussed, the latter novel 
insofar as the transfer rate to small scales is dominated by the 
helicity flux.

\section{Numerical simulations}
We solve numerically the equations for an incompressible rotating fluid
with constant mass density,
\begin{equation}
\frac{\partial {\bf u}}{\partial t} + \mbox{\boldmath $\omega$} \times
    {\bf u} + 2 \mbox{\boldmath $\Omega$} \times {\bf u}  =
    - \nabla {\cal P} + \nu \nabla^2 {\bf u} + {\bf F} ,
\label{eq:momentum}
\end{equation}
and
\begin{equation}
\nabla \cdot {\bf u} =0 ,
\label{eq:incompressible}
\end{equation}
where ${\bf u}$ is the velocity field,
$\mbox{\boldmath $\omega$} = \nabla \times {\bf u}$ is the vorticity,
${\cal P}$ is the total pressure (modified by the centrifugal term)
divided by the mass density, and $\nu$ is the kinematic viscosity.
Here, ${\bf F}$ is an external force that drives the turbulence, and
we choose the rotation axis to be in the $z$ direction:
$\mbox{\boldmath $\Omega$} = \Omega \hat{z}$, with $\Omega$ the
rotation frequency.

Eq. (\ref{eq:momentum}) is solved using a parallel pseudospectral
code in a three dimensional box of size $2\pi$ with periodic boundary
conditions and with a spatial resolution of $512^3$ regularly spaced 
grid points. The pressure is obtained by taking the divergence of Eq. 
(\ref{eq:momentum}), using the incompressibility condition 
(\ref{eq:incompressible}), and solving the resulting Poisson equation. 
The equations are evolved in time using a second order Runge-Kutta 
method, and the code uses the $2/3$-rule for dealiasing. As a result, 
the maximum wavenumber is $k_\textrm{max} = N/3$ where $N$ is the number 
of grid points in each direction. The code is fully parallelized with the 
message passing interface (MPI) library \cite{Gomez05a,Gomez05b}.

\begin{table}
\caption{\label{table:runs}Parameters used in the simulations. $k_F$ gives 
         the range of forcing wavenumbers, $\nu$ is the kinematic viscosity 
         and $\Omega$ the rotation rate; $Re$, $Ro$, and $Ek$ are respectively 
         the Reynolds, Rossby and Ekman numbers. Runs are performed on grids 
         of $512^3$ points in all cases and up to 40 turn-over times.}
\begin{ruledtabular} \begin{tabular}{ccccccc}
Run& $k_F$ &       $\nu$       &$\Omega$& $Re$  & $Ro$  & $Ek$              \\
\hline
A1 & 7--8  &$6.5\times 10^{-4}$& $0.06$ &$1200$ & $7.9$ & $6.5\times10^{-3}$ \\
A2 & 7--8  &$6.5\times 10^{-4}$&  $0.3$ &$1200$ & $1.6$ & $1\times10^{-3}$   \\
A3 & 7--8  &$6.5\times 10^{-4}$&   $7$  &$1200$ &$0.07$ & $6\times10^{-5}$   \\
A4 & 7--8  &$6.5\times 10^{-4}$&  $14$  &$1200$ &$0.03$ & $2.5\times10^{-5}$ \\
\hline
B1 & 2--3  & $6\times 10^{-4}$ & $0.08$ &$5700$ & $2.1$ & $4\times10^{-4}$   \\
B2 & 2--3  & $6\times 10^{-4}$ &  $3.5$ &$5700$ &$0.05$ & $9\times10^{-6}$   \\
B3 & 2--3  & $6\times 10^{-4}$ &   $8$  &$5700$ &$0.02$ & $3.5\times10^{-6}$ \\
\end{tabular} \end{ruledtabular}
\end{table}

The mechanical forcing ${\bf F}$ in Eq. (\ref{eq:momentum}) is given by 
the Arn'old-Beltrami-Childress (ABC) flow \cite{Childress}
{\setlength\arraycolsep{2pt}
\begin{eqnarray}
{\bf F} &=& F_0 \left\{ \left[B \cos(k_F y) + 
    C \sin(k_F z) \right] \hat{x} + \right. {} \nonumber \\
&& {} + \left[C \cos(k_F z) + A \sin(k_F x) \right] \hat{y} + 
   {} \nonumber \\
&& {} + \left. \left[A \cos(k_F x) + B \sin(k_F y) \right] 
   \hat{z} \right\},
\label{eq:ABC}
\end{eqnarray}}
\noindent where $F_0$ is the forcing amplitude, $A=0.9$, $B=1$, $C=1.1$ 
\cite{Archontis03}, and $k_F$ is the forcing wavenumber.  The ABC flow 
is an eigenfunction of the curl with eigenvalue $k_F$; as a result,
when used as a forcing function, it injects both energy and helicity in the 
flow. Table \ref{table:runs} gives the parameters used in the simulations. 
All runs are well resolved and were continued for over $40$ turnover times. 
Runs A1 and B1 were started from a fluid at rest, while the rest of the 
runs in sets A and B were started from the turbulent steady state reached 
at the end of runs A1 and B1 respectively.

The Reynolds, Rossby, and Ekman numbers are defined as usual as:
\begin{equation}
Re = \frac{L_F U}{\nu} ,
\end{equation}
\begin{equation}
Ro = \frac{U}{2 \Omega L_F} ,
\end{equation}
and
\begin{equation}
Ek = \frac{Ro}{Re} = \frac{\nu}{2 \Omega L_F^2} .
\end{equation}
where $L_F = 2\pi/k_F$, and the turnover time at the forcing scale is 
then defined as $T = L_F/U$ where $U=\left< u^2 \right>$ is the {\it r.m.s.}
velocity measured in the turbulent steady state or when the inverse 
cascade of energy starts (see below). The dissipation wavenumbers $k_\eta$ 
quoted below correspond to the Kolmogorov wavenumber 
$k_\eta = (\epsilon/\nu^3)^{1/4}$, where $\epsilon$ is the energy 
injection rate.

In the following, it will be useful to introduce a micro-Rossby number 
as the ratio of {\it r.m.s.} vorticity to the background vorticity 
(rotation), 
\begin{equation}
Ro_\omega = \frac{\omega}{2 \Omega} .
\label{eq:microRo}
\end{equation}
The value of the micro-Rossby number plays a central role in the inhibition 
of the energy cascade in rotating turbulence \cite{Cambon97}. If the 
micro-Rossby number is too small, non-linear interactions are completely
damped. According to \cite{Jacquin90}, anisotropies develop in rotating 
flows when the Rossby number $Ro \lesssim 1$ and when the micro-Rossby 
number $Ro_\omega \gtrsim 1$ (it is worth noting that the actual values 
for the transition depend on the particular flow studied).

The energy integral scale is given by
\begin{equation}
L = 2\pi \frac{\int_1^{k_\textrm{max}}{E(k) k^{-1} dk}} 
    {\int_1^{k_\textrm{max}}{E(k) dk}},
\label{eq:integral}
\end{equation}
where $E(k)$ is the isotropic energy spectrum (defined by averaging in 
Fourier space over spherical shells). An integral scale for the 
helicity can also be defined as
\begin{equation}
L^H = 2\pi \frac{\int_1^{k_\textrm{max}}{H(k) k^{-1} dk}}
    {\int_1^{k_\textrm{max}}{H(k) dk}}.
\label{eq:Hintegral}
\end{equation}
where $H(k)$ is the isotropic helicity spectrum. Perpendicular and 
parallel integral scales (e.g., $L_\perp$ and $L_\parallel$) are 
useful to measure the development of anisotropies and are defined by 
replacing $k$ by $k_\perp$ or $k_\parallel$ in Eqs. (\ref{eq:integral}) 
and (\ref{eq:Hintegral}). Here, the wavenumbers $k_\perp$ and $k_\parallel$ 
denote the reduced spectra -- e.g., $E(k_\perp)$ and $E(k_\parallel)$ -- 
computed averaging in Fourier space respectively over cylinders and over 
planes (see \cite{Mininni08} for more details). 

\section{Numerical Results}
\subsection{Energy inverse cascade at low Rossby numbers}

\begin{figure}
\includegraphics[width=8cm]{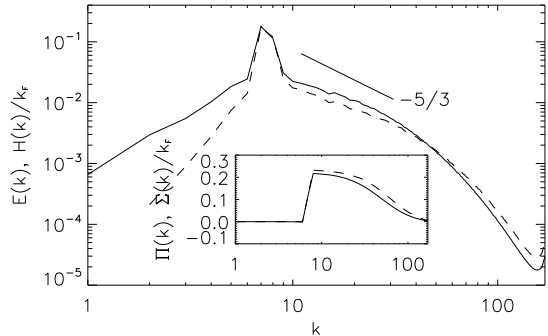}
\caption{Energy (solid line) and helicity (dash line) spectra in run A2 
    with forcing around $k\approx 7.5$ and almost negligible rotation. The 
    inset shows the energy and helicity fluxes indicative of two classical 
    direct cascades.}
\label{fig:spect20} \end{figure}

For strong rotation, it is known that the flow becomes quasi two-dimensional 
and an inverse cascade of energy is expected \cite{Cambon89,Waleffe93}. 
Figure \ref{fig:spect20} shows the energy and helicity spectra at late 
times in run A2, for a moderate Rossby number, as well as their fluxes. One 
observes that the flux of energy $\Pi(k)$ and of helicity $\Sigma(k)$ are 
both negligible for $k<k_F$ and are of order unity and positive at 
wavenumbers larger than the forcing wavenumber $k_F$ (here and in the 
following, the helicity spectrum and flux are plotted normalized by the 
forcing wavenumber $k_F$, to have them of the same order than the energy 
spectrum and flux when helicity injection is maximal). The inertial 
ranges of both the energy and helicity show similar scaling, close to 
K41 except for bottleneck (and possibly intermittency) corrections. Similar 
results are obtained in run A1 which has hardly any rotation effect.

\begin{figure}
\includegraphics[width=8cm]{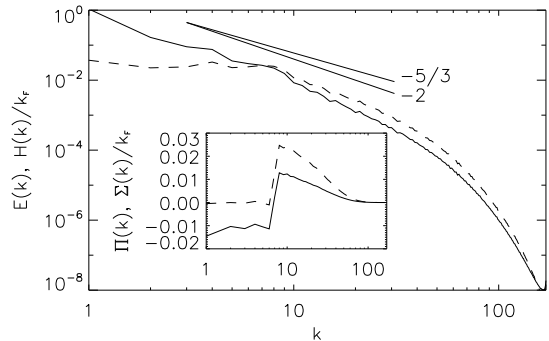}
\caption{Energy (solid) and helicity (dash) spectra in run A3 with same 
    forcing than run A2 but lower Rossby number. Different slopes are 
    shown as a reference. The inset gives the energy and helicity fluxes 
    and shows that there is both a direct and an inverse cascade of 
    energy but only a direct cascade of helicity.}
\label{fig:spect05} \end{figure}

However, runs A3 and A4 at low Rossby number show a different behavior (see 
Fig. \ref{fig:spect05}): at scales larger than the forcing scale, an inverse 
cascade of energy is observed, with constant and negative energy flux, and 
with its amplitude roughly an order of magnitude larger than in the large 
Rossby number case. However, the spectrum of helicity in this inverse range 
is approximately flat, and the flux of helicity towards large scales is 
almost negligible.

\begin{figure}
\includegraphics[width=8cm]{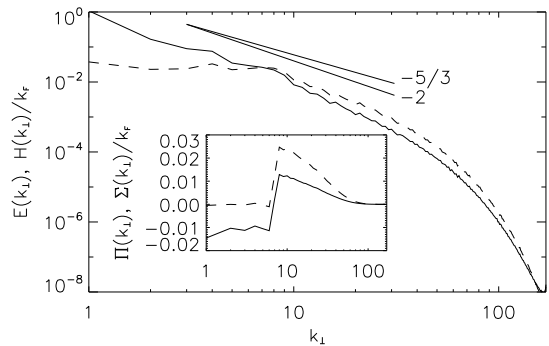}
\caption{Energy (solid) and helicity (dash) spectra as a function of 
    $k_\perp$ in run A3. Different slopes are shown as a reference. 
    The inset shows the energy and helicity fluxes in terms of $k_\perp$.}
\label{fig:sperp05} \end{figure}

The development of anisotropies and the inverse cascade of energy in 
rotating flows, leading for example to zonal flows in planetary atmospheres, 
has been explained in terms of near-resonant triad interactions of inertial 
waves: energy in three dimensional modes is transferred by a subset of the 
resonant interactions to modes with smaller vertical wavenumber 
\cite{Cambon89,Waleffe93}, a process that drives the flow to be quasi-two 
dimensional at large scales. The lack of an inverse transfer of helicity to 
large scales can be understood considering the partial two-dimensionalization 
of the flow at large scales: a helical flow is three-dimensional, while a 
two-dimensional flow has no helicity. Indeed, the energy spectra and fluxes 
in the direction perpendicular to $\mbox{\boldmath $\Omega$}$ are similar 
to the isotropic spectrum (see Fig. \ref{fig:sperp05}), while the spectrum 
in the direction parallel to $\mbox{\boldmath $\Omega$}$ peaks at 
$k_\parallel=0$ (details of how much energy is in the modes with 
$k_{\parallel}=0$ in each run are given in Table \ref{table:aniso}).

\begin{figure}
\includegraphics[width=8cm]{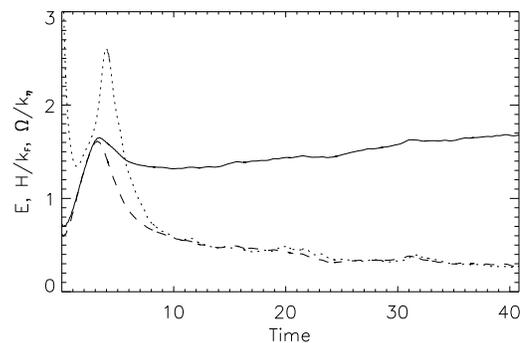}
\caption{Time evolution of the energy (solid), helicity (dash), and 
    enstrophy (dot) in run A3. The helicity is normalized by $k_F$, 
    and the enstrophy is rescaled by the dissipation wavenumber 
    $k_\eta \approx 100$. Only the energy undergoes an inverse 
    cascade, thereby growing with time.}
\label{fig:evolution} \end{figure}

The absence of an inverse cascade of helicity is further confirmed by 
the time evolution of the total energy, helicity and enstrophy (see 
Fig. \ref{fig:evolution}). While the energy increases monotonically after 
$t \approx 10$, the helicity and the enstrophy decay until reaching a 
steady state after $t \approx 25$. The monotonic increase of the total 
energy is the result of the piling up of energy at $k_\perp=1$ as the 
inverse cascade develops over time.

\begin{table}
\caption{\label{table:aniso}Anisotropies measured at $t \approx 40$ 
         in all runs . $Ro_\omega$ is the micro-Rossby number as 
         defined in Eq. (\ref{eq:microRo}), $L_\perp/L_\parallel$ is 
         the ratio of perpendicular to parallel integral scales as 
         defined in Eq. (\ref{eq:integral}), $L_\perp^H/L_\parallel^H$ 
         is the same ratio but based on the helicity spectrum as in 
         Eq. (\ref{eq:Hintegral}), $E(k_\parallel=0)/E$ is the ratio of 
         energy in all modes with $k_\parallel=0$ to the total energy, 
         and $H(k_\parallel=0)/H$ is the ratio of helicity in those 
         modes to the total helicity.}
\begin{ruledtabular}
\begin{tabular}{cccccc}
Run& $Ro_\omega$ & $L_\perp/L_\parallel$ & $L_\perp^H/L_\parallel^H$ & 
    $E(k_\parallel=0)/E$ & $H(k_\parallel=0)/H$ \\
\hline
A1 & $160$ & $0.56$ & $0.56$ & $0.05$ & $ 0.04$ \\
A2 &  $31$ & $0.55$ & $0.55$ & $0.06$ & $ 0.05$ \\
A3 & $0.6$ & $1.28$ & $0.53$ & $0.95$ & $ 0.74$ \\
A4 & $0.2$ & $1.27$ & $0.49$ & $0.98$ & $ 0.90$ \\
\hline
B1 &  $95$ & $0.86$ & $0.85$ & $0.30$ & $ 0.33$ \\
B2 & $1.1$ & $1.51$ & $1.20$ & $0.96$ & $ 0.85$ \\
B3 & $0.5$ & $1.36$ & $1.07$ & $0.96$ & $ 0.86$ \\
\end{tabular}
\end{ruledtabular}
\end{table}

However, the distributions of both  the energy and the helicity become 
anisotropic as time evolves. Table \ref{table:aniso} gives the micro-Rossby 
number for all the runs at $t \approx 40$, the ratios of perpendicular to 
parallel integral scales for the energy and for the helicity, 
$L_\perp/L_\parallel$ and $L_\perp^H/L_\parallel^H$, and finally
the amount of energy and helicity in the modes with $k_\parallel=0$ 
normalized respectively by the total energy and helicity. As the Rossby 
number decreases, the ratios $L_\perp/L_\parallel$ and 
$L_\perp^H/L_\parallel^H$ increase. However, the ratio of scales based 
on the helicity is smaller than the ratio of scales based on the energy, 
specially in the runs in set A where there is a larger separation 
between the largest scale in the box and the injection scale. This trend 
is accompanied by an increase in the amount of energy and helicity in the 
modes with $k_\parallel=0$, although here again the ratio 
$E(k_\parallel=0)/E$ is larger than $H(k_\parallel=0)/H$. This can be 
understood in terms of the Schwarz inequality for each mode in Fourier 
space. As the energy undergoes an inverse cascade, some helicity is 
transfered to the large scales (note the flat spectrum of helicity at 
large scales in Fig. \ref{fig:spect05} compared with the steep spectrum 
in Fig. \ref{fig:spect20}). According to the instability assumption of 
\cite{Waleffe93} (see also \cite{Greenspan69} and \cite{Cambon89}), 
the energy is transfered toward modes with wavevectors perpendicular to 
the rotation axis. From the Schwarz inequality, the helicity in each 
wave mode ${\bf k}$ must satisfy $|H({\bf k})| \le |{\bf k}| E({\bf k})$, 
and the large scale helicity must be transfered towards $k_\perp$ to 
satisfy this relation.

\subsection{The case for direct cascades}
\begin{figure}
\includegraphics[width=8cm]{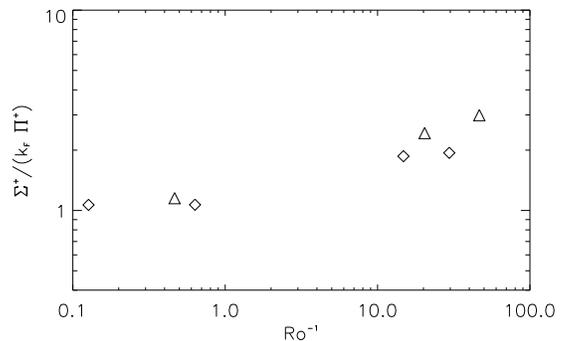}
\caption{Ratio of helicity flux to energy flux towards small scales as a 
    function of inverse Rossby number and at a fixed Reynolds number for 
    each set (see Table I). Diamonds correspond to runs in set A, and 
    triangles correspond to runs in set B. Note the increase in relative 
    strength of the helicity cascade to small scales as rotation increases.}
\label{fig:ratio} \end{figure}

At scales smaller than the forcing scale, the energy spectrum in runs A3 
and A4 at low Rossby numbers is slightly steeper than $k^{-2}$ (see Fig. 
\ref{fig:sperp05}), and (unlike the case of non-rotating turbulence), the 
helicity spectrum is possibly shallower than the energy spectrum (a 
confirmation of this using runs in set B is discussed below). Furthermore 
(see Fig. \ref{fig:ratio}), the energy flux $\Pi(k)$ becomes smaller than 
the (normalized) helicity flux $\Sigma(k)/k_F$ at wavenumbers larger than 
$k_F$ as the Rossby number is decreased.

This change can be understood as follows. The energy injection rate 
$\epsilon$ and the helicity injection rate $\delta$ are related by 
$\delta \sim  k_F \, \epsilon$ (these two quantities are equal when 
maximally helical forcing is applied at a single wavenumber). The 
Schwarz inequality in each shell $|H(k)|\le kE(k)$ implies that, at 
large scale (in the limit $k\rightarrow 0$), there must be a negligible 
flux of helicity (unless of course $E(k) \rightarrow \infty$); thus 
helicity is bound to cascade to small scales. However, the development 
of an inverse cascade of energy decreases the amount of energy flux 
that can go to small scales, and as a result the helicity flux 
dominates for $k>k_F$. This can be illustrated by plotting the ratio 
$\Sigma^+/(k_F \Pi^+)$ (Fig. \ref{fig:ratio}), where $\Sigma^+$ and 
$\Pi^+$ denote respectively the amount of helicity and energy flux 
that goes towards small scales. Note that $\Sigma^+/(k_F \Pi^+) \approx 1$ 
for $Ro > 1$ (both quantities direct cascade), while as the Rossby 
number decreases $\Sigma^+/k_F > \Pi^+$.

\begin{figure}[h!]
\includegraphics[width=8cm]{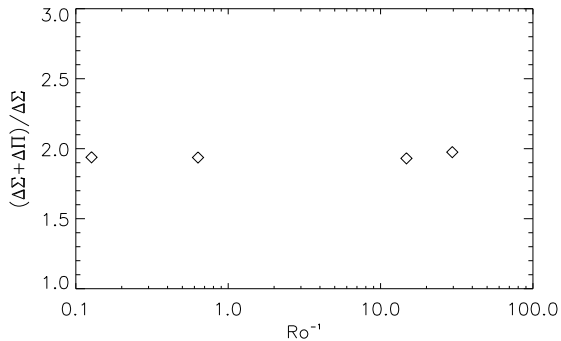}
\caption{Sum, for the energy and helicity, of the normalized differences 
    between their direct and inverse fluxes (see text). Only runs in set 
    A (diamonds) are shown because runs in set B do not have enough scale 
    separation between the forcing and the largest scale in the box to 
    compute $\Sigma^-$ and $\Pi^-$ reliably.}
\label{fig:diff} \end{figure}

We can also introduce the differences between the direct and inverse
energy and helicity fluxes, respectively as 
$\Delta \Sigma = (\Sigma^+ - \Sigma^-)/k_F$ and $\Delta \Pi = \Pi^+ - \Pi^-$
(where $\Sigma^-$ and $\Pi^-$ are negative and denote respectively the 
amount of helicity and energy that go towards large scales). Figure 
\ref{fig:diff} shows the normalized ratio 
$$\rho = (\Delta \Sigma + \Delta \Pi)\ /\ \Delta \Sigma \ .$$
This ratio is roughly independent of the Rossby number, which further 
confirms that the dominance of the helicity flux for $k>k_F$ is associated 
with the energy flux lost in that range because of the inverse cascade 
of energy.

As a result, the direct transfer in the small scales of a rotating helical 
turbulent flow is dominated by the (normalized) helicity flux. In the limit 
of a pure helicity cascade with no direct energy cascade, and considering 
the effect of rotation, the helicity flux can be expressed as 
\begin{equation}
\Sigma(k) \sim \delta \sim \frac{h_\ell}{\tau_\ell^2} \tau_\Omega ,
\label{eq:flux}
\end{equation}
where $h_\ell$ is the helicity at the scale $\ell$, 
$\tau_\ell \sim \ell/u_\ell$ is the eddy turnover time at the same 
scale, and $\tau_\Omega \sim 1/\Omega$ is the characteristic time of 
inertial waves. This expression takes into account the slowing-down of 
transfer to small scales due to three-wave interactions (see e.g., 
\cite{Zhou95,Muller07}), in a similar fashion as what was proposed by 
Iroshnikov and Kraichnan for Alfv\'en waves in the presence of a magnetic 
field \cite{Iroshnikov63,Kraichnan65} (the extension to the anisotropic 
case can be trivially obtained considering the turnover time as 
$\tau_\ell \sim \ell_\perp/u_\ell$, see e.g., \cite{Muller07}). From 
this expression, it follows that if 
$E(k) \sim k^{-n}$ ($n \le 2.5$, with the equality holding for the case 
with maximum helicity at all scales from Schwarz inequality), then
\begin{equation}
H(k) \sim k^{n-4} ,
\label{eq:spectrum}
\end{equation}
i.e., resulting in a shallower helicity spectrum for $n\ge 2$ (note that 
for $n<2$ the helicity spectrum is steeper than the energy spectrum).

\begin{figure}
\includegraphics[width=8cm]{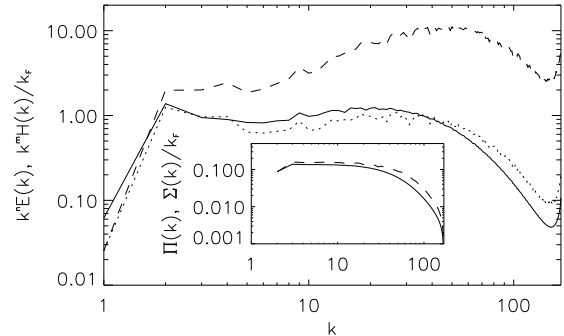}
\caption{Energy spectrum compensated by $k^n$ with $n=5/3$ (solid), 
    helicity spectrum compensated by $k^m$ with $m=n$ (dots), and 
    compensated by $m=n-4 \approx 2.33$ (dash line), in run B1 
    with large-scale forcing and weak rotation. Note that the helicity 
    and the energy in this run have the same Kolmogorov scaling in 
    the inertial range. The inset shows the energy and normalized 
    helicity fluxes with solid and dash lines respectively.}
\label{fig:spect20b} \end{figure}

\begin{figure}
\includegraphics[width=8cm]{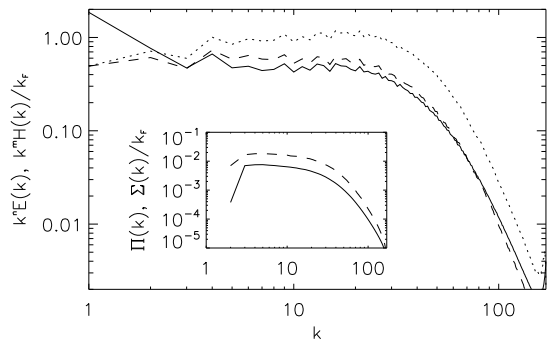}
\caption{Energy spectrum compensated by $k^n$ with $n=2.15$ (solid), 
    helicity spectrum compensated by $k^m$ with $m=n$ (dot), and 
    compensated by $m=n-4 = 1.85$ (dash line), in run B2 with 
    large-scale forcing and low Rossby number. Note that the helicity 
    and the energy spectra in this run have different scalings in the 
    inertial range (different from each other and different from K41), 
    and both are flat only when compensated following Eq. (\ref{eq:spectrum}). 
    The inset again shows the energy and (normalized) helicity fluxes; 
    note the domination of the latter in this high-rotation regime.}
\label{fig:spect05b} \end{figure}

Although the runs in set A have a helicity spectrum that is indeed
slightly shallower than the energy spectrum, the forcing is applied at 
intermediate scales and the scale separation between the forcing and 
dissipative scales is not enough to confirm the scaling prediction of Eq. 
(\ref{eq:spectrum}). Indeed, the micro-Rossby numbers are 
$Ro_\omega \approx 0.6$ for run A3 and $\approx 0.2$ for run A4 (see 
Table \ref{table:aniso}). A larger direct inertial range and larger 
micro-Rossby numbers are needed in order to check the validity of the 
predicted scaling.

To that effect, we now report on the runs in set B (see Tables 
\ref{table:runs} and \ref{table:aniso}) which have a forcing function 
concentrated in the large scales; the inverse cascade is thus not so well 
resolved but it allows for a more developed direct inertial range. In 
particular, we will focus on run B2 which as a Rossby number $Ro\approx 0.05$ 
and a micro-Rossby number $Ro_\omega \approx 1.1$. Figures \ref{fig:spect20b} 
and \ref{fig:spect05b} show the compensated energy and helicity spectra 
for runs B1 and B2 (run B3 behaves as run A3). It is observed that while 
in run B1 (corresponding to weak rotation), the helicity and energy 
spectra have the same scaling ($\sim k^{-5/3}$ with bottleneck and 
intermittency corrections), in run B2 the compensated helicity and 
energy spectra are horizontal and parallel only when using the scaling 
law predicted by Eq. (\ref{eq:spectrum}).

The same scaling is observed in $k_\perp$. As previously mentioned, it 
is straightforward to recast Eqs. (\ref{eq:flux}) and (\ref{eq:spectrum}) 
to take into account the anisotropies in the flow, again similarly to the 
magnetohydrodynamic case. The results exemplified by Fig. 
\ref{fig:spect05b} confirm that the small-scale scaling of energy and 
helicity differ in rotating turbulence, unlike the non-rotating case 
(see e.g., Fig. \ref{fig:spect20b}) where energy and helicity follow 
the same spectral laws.

\section{Conclusion}

Even though the Rossby number in the atmosphere of the Earth is not very 
large, the existence of inertial waves that can interact with turbulent 
eddies is bound to affect the dynamics of the turbulent flow, as has been 
studied by several authors. Helicity, which is also observed in atmospheric 
flows, is known to play an important role in the evolution of tornadoes. 
But, as already found in \cite{Morinishi01}, the two physical phenomena 
(rotation on the one hand, helicity on the other hand) reduce nonlinear 
interactions in different ways. Thus a study combining both effects at 
high Reynolds number can shed some light on the dynamics of such flows. 

This paper shows that for strong rotation, the direct cascade to small 
scales is now dominated by the helicity flux (and the inverse cascade, as 
expected, by the energy). Moreover, the resulting spectrum is different 
from what Kolmogorov scaling predicts for the non-rotating case, and 
from what a pure direct cascade of energy slowed down by eddy-wave 
interactions predicts for the rotating non-helical case. In this context 
it is worth mentioning that, using phenomenological arguments, a direct 
cascade of helicity in rotating flows has also been argued recently 
in \cite{Chakraborty07}, although the arguments predicted a different 
scaling and were based on Fjortoft's theorem which does not necessarily 
apply to the helicity since is not a positive definite quantity 
\cite{Kraichnan73}.

A novel phenomenological argument based on a cascade of helicity towards 
small scales slowed down by wave-eddy interactions lead to different 
inertial indices for the small-scale energy and helicity spectra, and 
provides a good fit to the results of the simulations presented in this 
paper. The spectral indices are bounded by the value that corresponds 
to a flow with maximum helicity, and depend on the amount of relative 
helicity in the flow. The result differs from non-rotating turbulence, 
where the energy and the helicity follow the same scaling laws 
\cite{Brissaud73,Chen03,Gomez04}. Although the DNS runs confirm the 
scaling, due to computational limitations well-resolved inverse and 
direct cascades had to be studied in separate simulations. In the 
future, a simulation of helical rotating turbulence at very large 
resolution will be performed to confirm these results with a better 
resolved coexistence of the direct and inverse cascades.

The study of the intermittency of a mixture of turbulence and waves in 
the presence of rotation and helicity will also be the topic of a future 
work; it is of particular interest since it will shed some light on the 
statistics, structures and interactions of extreme events which, when 
combined with realistic physics of the atmosphere (e.g., adding weak 
compressibility, moisture and geometry), will lead eventually to a better 
understanding and prediction of the behavior of atmospheric flows.

\begin{acknowledgments}
Computer time was provided by NCAR. NCAR is sponsored by the National 
Science Foundation. PDM acknowledges support from grant UBACYT X468/08 
and is a member of the Carrera del Investigador Cient\'{\i}fico of CONICET.
\end{acknowledgments}

\bibliography{ms}

\end{document}